     \documentstyle[12pt,agums1]{article}

\lefthead{ANTIOCHOS AND DEVORE}

\righthead{MAGNETIC RECONNECTION AND SOLAR ACTIVITY}

\authoraddr{S. K. Antiochos,
Code 7675,
Naval Research Lab,
Washington, DC 20375-5352.
(e-mail: antiochos@nrl.navy.mil}

\authoraddr{C. R. DeVore,
Code 6440,
Naval Research Lab,
Washington, DC 20375.
(e-mail: devore@lcp.nrl.navy.mil)}

\begin{document}

%
%

\title{The Role of Magnetic Reconnection in Solar Activity}

%
%

\author{Spiro K. Antiochos and C. Richard DeVore}
\affil{Naval Research Laboratory, Washington, D. C.}

%
%

\begin{abstract}
We argue that magnetic reconnection plays the determining role in many
of the various manifestations of solar activity. In particular, it is
the trigger mechanism for the most energetic of solar events, coronal
mass ejections and eruptive flares. We propose that in order to obtain
explosive eruptions, magnetic reconnection in the corona must have an
``on-off'' nature, and show that reconnection in a sheared multi-polar
field configuration does have this property. Numerical simulation
results which support this model are presented, and implications for
coronal mass ejections/eruptive flare prediction are discussed.
\end{abstract}


\begin{article}
\section{INTRODUCTION}

Magnetic reconnection is widely believed to be the most important
process by which magnetic fields transfer energy to plasmas and,
consequently, has been invoked as the explanation for a multitude of
phenomena observed in space, astrophysical, and laboratory plasmas.
Reconnection is the most likely mechanism by which the Sun's magnetic
field powers the solar atmosphere, the chromosphere and
corona. Reconnection is also the process by which the solar wind flow
couples to planetary magnetospheres, and is the key process behind the
eventual transfer of this energy to the Earth's ionosphere and lower
atmosphere. Hence, it would not be unjustified to describe the ISTP as
a program to study the effects of reconnection on the Sun-Earth
Connection.

Reconnection is particularly attractive as an explanation for solar
variability and activity because of the potential for both rapid
dissipation of magnetic energy and the accompanying strong plasma
dynamics. In general, one can categorize the energy release due to
reconnection as taking four distinct forms, all of which are believed
to occur commonly in the solar atmosphere.  First, reconnection can
result in direct plasma heating. This has been proposed as the process
that heats both the non-flare 
[e.g., \markcite{{\it Parker,} 1972, 1979, 1983}; 
\markcite{{\it Sturrock and Uchida,} 1981}; 
\markcite{{\it van Ballegooijen,} 1986}; 
\markcite{{\it Antiochos,} 1990}] and flare
corona [e.g., \markcite{{\it Sweet,} 1958}; 
\markcite{{\it Parker,} 1963}; \markcite{{\it Petschek,} 1964}; 
\markcite{{\it Carmichael,} 1964}; \markcite{{\it Sturrock,} 1966}] 
Second, reconnection can produce strong mass motions,
the so-called reconnection jets. These have been proposed 
[\markcite{{\it Karpen, Antiochos, and DeVore,} 1995, 1996, 1998}] 
as the explanation for a large variety of
transient dynamic phenomena ranging from spicules 
[e.g., \markcite{{\it Blake and Sturrock,} 1985}] 
to explosive events 
[\markcite{{\it Dere et al.,} 1991}] 
to large surges and sprays 
[e.g.,  \markcite{{\it Rust,} 1968}; \markcite{{\it Herant et al.,} 1991}; 
\markcite{{\it Schmeider et al.,} 1994};
\markcite{{\it Yokohama and Shibata,} 1996}] 
and X-ray jets [\markcite{{\it Shibata et al.,} 1992}]. Third,
reconnection can lead to the acceleration of non-thermal
particles. This process has been proposed as the explanation for the
electron beams that give rise to the hard X-ray emission in flares
[e.g., \markcite{{\it Sturrock,} 1980 {\it and references therein}}]. 
Finally, reconnection can
produce MHD waves. This process has been proposed as the explanation
for coronal heating [e.g., \markcite{{\it Falconer et al.,} 1997}] 
and for accelerating the solar wind 
[e.g., \markcite{{\it Parker,} 1988}; \markcite{{\it Mullan,} 1990}]. 
In fact, it is difficult to find a coronal phenomena that has not 
been attributed to magnetic reconnection!

Rather than considering one of these direct effects of reconnection,
we discuss in this paper one of the most interesting and more subtle
roles of reconnection --- the initiation of major disruptions of the
Sun's magnetic field.  Large magnetic disruptions are generally
observed as a coronal mass ejection (CME) and a flare associated
prominence/filament eruption.  We will refer to them as CME/eruptive
flare (EF). These events are the most energetic (up to $10^{33}$ ergs)
and destructive of solar disturbances. They are the main drivers of
space weather and are a focus of study of the ISTP program. The May 2
1998 event, for example, was observed by SOHO/ISTP to consist of a
filament eruption, fast CME (1,000 km/s) and an X-class flare that
produced a strong proton storm at Earth and the strongest geomagnetic
activity in recent years ($K_p = 9$). CME/EFs will be of particular
interest in the upcoming maximum.

In addition to their key role in driving space weather, CME/EFs are
important from the viewpoint of understanding basic space-plasma
physics. They are the classic manifestation of solar activity, and the
mechanism for their initiation and evolution has long been a central
issue of solar physics research [e.g., \markcite{{\it Sturrock,}
1980}]. Space observations such as those from Yohkoh and SOHO have led
to much progress in understanding magnetic disruptions, and a general
framework for these events has emerged.  A CME/EF is believed to
consist of four steps: {\bf (1.)}\ Stressed magnetic field slowly
emerges from below the photosphere building up the free energy in the
corona quasi-statically.  The emerged field may be further stressed by
the convective and rotational photospheric motions. {\bf (2.)}\ Some
triggering mechanism destroys the magnetic equilibrium of the corona,
resulting in an explosive instability or loss of equilibrium.  {\bf
(3.)}\ The magnetic field erupts outward, accelerating and ejecting
plasma into the heliosphere. {\bf (4.)}\ The field closes back down
via reconnection to a less stressed state, heating the plasma and
producing the intense X-ray burst.

In this four-step picture, a CME/EF can be thought of as simply the
method by which the Sun rids itself of magnetic stress. Since the coronal
conductivity is so high, the Sun has no choice but to eject any
large-scale magnetic stress into the heliosphere.  Note also that in
this picture, a CME and an eruptive flare are part of a single
phenomenon. The CME and prominence/filament eruption constitute step
{\bf (3.)}, the opening of the field.  It should be emphasized that by
opening we don't mean that the field actually disconnects from the
Sun, only that the field becomes dynamic and expands outward without
limit.  During this step the free magnetic energy is used primarily to
accelerate plasma and lift it against gravity. The X-ray or
H$_\alpha$ flare constitutes step {\bf (4.)}, the closing of the field.
During this step the energy is used primarily to heat magnetically
confined plasma and to accelerate nonthermal particles.

As with any astrophysical phenomenon, one can always find CMEs and
flares that appear to be exceptions to the standard picture described
above. Many CMEs, especially high-latitude slow ones, have no
detectable X-ray emission.  In this case the reconnection during step
{\bf (4.)} is presumably very slow, resulting in minimal heating. Much
more difficult to reconcile, however, is the observation that intense
X-ray flares sometimes occur with no apparent magnetic field opening
i.e., no CME or prominence/filament eruption.  Strong flare heating is
widely believed to be due to fast reconnection, which generally involves
the formation of a current sheet.  It is the opening of the field
during {\bf (3.)} that forms a large current sheet, thereby, allowing
rapid reconnection to occur during {\bf (4.)}.  In order to account
for non-eruptive flares, it must be that either the field does erupt
but produces undetectable plasma motions --- this seems unlikely ---
or else large current sheets can form and fast reconnection can occur
even in fully closed bipolar configurations. There are good arguments,
however, against the formation of current sheets in such a magnetic
topology [e.g., \markcite{{\it van Ballejooigen,} 1985}; 
\markcite{{\it Antiochos,} 1987}], furthermore
reconnection is generally believed to be inhibited by line-tying
effects for such topologies [\markcite{{\it Dahlburg, Antiochos, and Zang,}
1991}; \markcite{{\it Velli, Einaudi, and Hood,} 1993}].
More observational and theoretical work is clearly needed on the
issue of apparently non-eruptive flares.

In this paper we adopt the standard 4-step picture above for a CME/EF
and focus on Step {\bf (2.)}, the triggering mechanism.  Determining
the flare trigger has long been one of the outstanding problems in
solar physics [e.g., \markcite{{\it Sturrock,} 1980}].  In addition,
it is clearly necessary to understand the triggering process if we are
ever to develop physics-based CME and eruptive flare prediction
schemes. We argue below that magnetic reconnection in a multi-flux
magnetic topology is this long-sought-after triggering mechanism.

\section{THEORETICAL RESULTS}

Over the years many models have been proposed for the initiation of
CMEs and EFs, the large majority of which are magnetically driven
[e.g., \markcite{{\it Sturrock,} 1989}; \markcite{{\it Moore 
and Roumeliotis,} 1992}; \markcite{{\it Mikic and Linker,}
1994}; \markcite{{\it Antiochos, DeVore, and Klimchuk,} 1998}],
although models in which gas buoyancy
effects are important [e.g., \markcite{{\it Low,} 1994}; 
\markcite{{\it Wolfson and Dlamini,} 1997}] have
also been discussed. Since the solar corona is generally measured to
have a low plasma beta $< 10^{-2}$, it is almost certain that the
energy for the eruption is stored in the magnetic field. This is
especially true for the very energetic, fast eruptions which can have
speeds exceeding 2000 km/s -- of order the Alfven speed, far larger
than coronal sound speeds.

In recent years, however, a major obstacle to the magnetic models has
been pointed out by Aly and Sturrock. Their argument proceeds as
follows. Aly (1984) in a landmark paper proved that the energy of any
force-free magnetic field configuration in the corona is bounded
above. This result implies that if the coronal field is stressed by
photospheric motions, the energy of the field will not continue rising
indefinitely even if the motions continue indefinitely. Aly showed
that, in fact, the energy has an upper bound that is only of order the
energy of the potential magnetic field with the same normal flux
distribution at the photosphere.  For example, a dipole field is
bounded above by twice the potential field energy [\markcite{{\it
Aly,} 1984}].  The physical reason for the energy bound is
straightforward.  The corona has an infinite volume, and so if even
the footpoint motions become infinitely large, the field can simply
expand outward toward infinity, thereby, keeping its energy finite.
Note that this argument is equally valid if the magnetic stress is due
to emergence from below rather than footpoint motions.  As long as the
evolution is quasi-static, the magnetic free energy and also the
electric current, or magnetic stress, in the corona must have a strict
upper bound [e.g., \markcite{{\it Aly,} 1984}; \markcite{{\it Finn and
Chen,} 1990}].  Therefore, one must be careful in specifying boundary
conditions for the corona at its photospheric base.  For example, one
is not free to specify arbitrarily the current entering the corona
from the photospheric. In general, the safest and most physical
boundary conditions are to specify footpoint motions, which is our
approach in the simulations described below.

Building upon this work, Aly (1991) presented arguments that if the
coronal magnetic field undergoes a quasi-static evolution, which is a
good approximation for photospheric driving motions, then the least
upper bound on its energy is the open field configuration in which
every field line is stretched out to infinity. Sturrock (1991)
independently presented similar arguments. Although their arguments do
not constitute a rigorous proof, they are quite convincing and appear
to be in good agreement with every numerical simulation performed to
date [e.g., \markcite{{\it Roumeliotis, Antiochos, and Sturrock,}
1994}; \markcite{{\it Mikic and Linker,} 1994}; \markcite{{\it Amari
et al.,} 1996}], (but see \markcite{Wolfson and Low} (1992) for a
possible counter example). The Aly-Sturrock energy limit also seems
physically intuitive. The basic assumption is that the coronal field
is always free to lower its energy by an ideal outward expansion, if
such an expansion is energetically favorable. Consequently, it should
not be possible to build up quasi-statically significant energy above
the fully open state, because as soon as the magnetic energy tries to
exceed the open value, the field would simply expand to the open
state. Once the field is open, further footpoint motions have no
effect other than to propagate Alfven waves out into the heliosphere.

Although the energy limit seems perfectly reasonable, it also seems
completely incompatible with CME/EF observations. If these events are
magnetically driven, the energy in the pre-eruption magnetic field
must be considerably larger than the open field state formed during
the eruption, contrary to the energy limit.  It is often claimed that
one can circumvent the energy limit by simply opening part of the
field. For example, one can leave the very low-lying field closed and
open only above a certain height. We believe that this argument is
spurious. While it is true that in their papers Aly and Sturrock
concentrated on the fully open field case, as we discussed above,
their result is based on quite general arguments that should hold even
for a partial opening of the field.  The basic point is that a slowly
driven magnetic field is unlikely to erupt explosively to an open
state that it can also reach by a process of ideal expansion.  Indeed,
EIT and LASCO/C1 observations from SOHO show that slow expansion is a
general feature of the coronal magnetic field.  Furthermore, we
believe that most CMEs, which are typically slow and accelerate only
up to slow-solar wind speeds, also represent merely a slow expansion
of the field.  On the other hand, there is no doubt that explosive
eruptions do sometimes occur.

\section{NUMERICAL RESULTS}

An attractive solution to this apparent contradiction between theory
and observation is to invoke reconnection. It seems likely that the
extra freedom in the possible evolution of the field afforded by
reconnection can somehow circumvent the energy limit. This is exactly
what happens in 2.5D geometries. Mikic and Linker (1994) showed that
footpoint shearing in a system with azimuthal symmetry and finite
resistivity leads to the formation of disconnected plasmoids that are
expelled rapidly from the corona with a significant energy drop. We
emphasize again that, contrary to common misconception, the eruption in
this case is not due the fact that the field opens only partially.
The eruption occurs because the field opens to a state that it cannot
access by an ideal expansion.  The plasmoid carries off a large
fraction of the shear of the closed field lines, so that the final
state of the system is one in which the closed region contains much
less shear than was originally imparted to this region by the
footpoint motions. The field could not erupt to the state in which
the shear in the closed field region stayed fixed.

This discussion leads to the main conclusion of this paper: the whole
eruption process is controlled by the nature of magnetic reconnection
in the corona.  In order for plasmoid formation to result in an
explosive eruption, the formation and hence the reconnection must
proceed rapidly. The plasmoid must grow faster than the field expands
outward.  On the other hand, reconnection must not occur too easily,
otherwise very little free energy could be built up. Therefore, to
explain eruptions, magnetic reconnection in the solar corona must have
a very particular form. It must have an ``on-off'' nature wherein it
is basically off during step {\bf (1.)} while the free energy is
building up, but at a certain point switches on to a high value {\it
and stays on}. It is this reconnection switch-on that is the flare
trigger, step {\bf (2.)}.

Since plasmoid formation is only a 2.5D phenomenon, it is not clear,
however, whether reconnection can also lead to an eruption in a full
3D system.  The 3D analogue to the 2.5D configuration is the so-called
tether-cutting model [\markcite{{\it Sturrock,} 1989}; \markcite{{\it
Moore and Roumeliotis,} 1991}].  The basic geometry of this model can
be seen in Plate 1, which shows the results of a 3D MHD simulation of
photospheric stressing of a coronal field.  This simulation is a fully
time-dependent calculation of our model for prominence formation
[\markcite{{\it Antiochos, Dahlburg, and Klimchuk,} 1994};
\markcite{{\it Antiochos,} 1995}], and corresponds to a 3D
generalization of the 2.5D CME calculations.  The magnetic field in
the figure is the result of applying a shear flow at the photosphere
to an initially current-free dipole field. The plane of the
photosphere in this simulation was taken to be a rectangle, part of
which can be seen in the figure. The straight black line running
lengthwise through the middle of the photospheric plane is the
magnetic polarity reversal line. In agreement with observations
[\markcite{{\it Schmeider et al.,} 1996}], the shear was chosen to be
spatially localized near the polarity reversal line.  The form of this
shear can be seen in the contours of the normal component of $B$ on
the photospheric plane; for ease of viewing only one contour is shown
on the far side of the bottom plane. Note that in the corona above
this plane there are basically two types of field lines. Those
originating in the shear zone (three thick lines) are stretched out
and have concave up portions along their length, while those
originating in the non-shear zone (three thin arrowed lines) appear to
be dipolar. The mass of red lines in the center of the figure indicates
where the magnetic field lines are concave up and represents the 
H$_\alpha$ mass of the prominence.

The idea behind tether-cutting is that reconnection leads to eruption
by redistributing the shear so that it becomes concentrated on the
furthermost field lines at the edge of the sheared region. The field
lines around the center of the figure decrease their shear, while
those near the edges increase.  Such a transfer of shear to the
outermost field lines is the physical analogue to plasmoid formation
in 2.5D.  If this shear transfer is sufficiently rapid, then eruption
should occur.  Hence, the 3D case is really not significantly different
than the 2.5D system. In both cases the question of whether an
eruption occurs comes down to the issue of the temporal behavior of
reconnection in the solar corona

So far, our 3D simulations of tether-cutting show no evidence for an
explosive eruption, which agrees with our previous work on the
twisting of dipolar arcades [\markcite{{\it Dahlburg, Antiochos, and
Zang,} 1991}].  However, this work
is still in progress, and it may be that using different parameters,
such as system size, form of shear, etc. will lead to
eruption. Further studies of the tether-cutting model are needed
before definitive conclusions can be reached.

Based on our work so far, however, we conjecture that the
tether-cutting model does not lead to explosive eruptions. The basic
reason for this result is that a bipolar field system as in Plate 1 is
too simple topologically for fast reconnection to occur. We propose
instead the ``breakout'' model [\markcite{{\it Antiochos,} 1998};
\markcite{{it Antiochos, DeVore, and Klimchuk,} 1999}] shown in Plate
2, which consists of a multi-polar field. The initial current-free
field for this system is given by the vector potential:
\begin{equation}
\vec{A} = {\sin\theta \over r^2} + 
{(3 + 5\cos 2\theta)\sin\theta \over 2 r^4} \; \hat\phi
\end{equation}
This system has a more complex topology than that of Plate 1, it
consists of four flux systems with separatrix surfaces and a null
point in the corona, as shown in the first panel of Plate 2.  There
are now three neutral lines in the system, at solar latitudes of $\pm
45^\odot$ and at the equator. Note, however, that this field
configuration arises from only four polarity regions on the
photospheric surface, hence it is actually far simpler than what must
be present in the real corona. Also it appears similar to what is
often seen in the LASCO C1 images.

The effect of stressing this system by footpoint motions can be seen
in Plate 2, which present the results from one of our 2.5D MHD
simulations. For this case we selected the inner flux system
(straddling the equator), as the one to shear.  Again the shear is
concentrated near the neutral line. The thick dark lines in Plate 2
correspond to the sheared prominence field lines of Plate 1.
Basically, we have embedded the sheared arcade of Plate 1 into a
more global field topology.  As a result of being sheared, the inner
flux system overlying the equator expands outward, just as the single
arcade in Plate 1 expands outward. But now it begins to interact with
the neighboring systems. As a result of this expansion the null point
in the corona deforms into a neutral sheet, similar to the process
originally proposed by Syrovatskii (1981). Eventually, this neutral
sheet becomes sufficiently thin that reconnection begins. We see in
the last panel of Plate 2 that some of the field lines that used to be
in the inner and outer flux systems have transferred over to the
sides.  Once this reconnection starts it continues to accelerate
because there are fewer unsheared field lines holding down the sheared
ones.  The rapidly accelerating transfer of flux from above the
sheared field allows this field to ``break out'' and erupt open.

Note that our breakout model yields a very natural process for
obtaining an ``on-off'' type of reconnection in the corona.  In fact,
we find that once the reconnection switches on, the rate becomes too
large for our ideal code to simulate, but the essential result has
been shown -- in a multi-flux system reconnection will have the
correct time-dependence to explain the initiation of CME/EFs, step
{\bf (2.)}. Hence, we conclude that reconnection is responsible not
only for releasing a large fraction of the flare energy during step
{\bf (4.)}, it is also responsible for getting the whole event going.

Now that we have a likely candidate for the trigger mechanismm, it is
interesting to consider its use as a possible CME/EF predictor. There
are two distinguishing features of the breakout model. One is that a
multi-polar configuration is needed for eruption. In fact, it is well
known that complexity is needed to obtain eruptive flares, and
magnetic complexity is already one of the main criteria being used for
prediction. The second feature is that reconnection should occur high
up in the corona before the eruption. We believe that it would be
highly informative to use the resources of the ISTP to search for this
reconnection during the upcoming solar maximum.


\acknowledgments
This work was supported in part by NASA and ONR.


%
%

\end{article}

\newpage

	\begin{plate} 
\platenum{1} 
\platewidth{41pc} 
\caption{Results from a 3D MHD simulation of our model for
prominence/filament formation and of the tether cutting model
for coronal mass ejections/ eruptive flares. The
configuration shown is that of an initial current-free dipole field
that has been sheared by photospheric motions.  The outline of the
bottom boundary of our simulation box is shown along with the
photospheric neutral line. Contours of normal-magnetic-field magnitude
are plotted on the bottom boundary plane.  Three field lines (thick
dark-blue) with footpoints in the shear region and three field lines
(arrowed, light-blue) with footpoints outside this region are
shown. Also plotted are contours of positive-upward curvature of
magnetic field (mass of red lines), which corresponds to the possible
location of the prominence mass.}
\end{plate}
	\begin{plate}
\platenum{2}
\platewidth{41pc}
\caption{Results from a 2.5D simulation of the breakout model for
coronal mass ejections and eruptive flares. The first panel (upper
left) shows the initial current-free configuration which consists of a
sum of a dipolar and an octopolar field. Photospheric shearing is
applied only near the equatorial neutral line. The next three panels
show the field after 17.5, 19.7, and 21.7 hours into the evolution.
All field lines are traced from exactly the same footpoint positions
in all panels. Note that in the last panel the 
overlying unsheared field is reconnecting and moving to the side,
thereby allowing the sheared flux to erupt outward.}
\end{plate}

\end{document}